\documentstyle[pre,aps,multicol]{revtex}
\input{epsf}

\begin{document}
\draft
\title{Convection in rotating annuli:
Ginzburg-Landau equations with tunable coefficients}
\author{Martin van Hecke\footnote{Address after Oct, 1 1996: The Niels Bohr Institute, 
Blegdamsvej 17, 
2100 Copenhagen \O, Denmark} 
and Wim van Saarloos}
\address{Instituut-Lorentz, Leiden University,
P.O. Box 9506, 2300 RA Leiden, the Netherlands.}

\date{\today}
\maketitle
\begin{abstract}
The coefficients of the 
complex Ginzburg-Landau equations  that 
describe weakly nonlinear convection in a large 
rotating annulus are calculated for a range of Prandtl numbers
$\sigma$.  
For fluids with 
$\sigma \approx 0.15$, we show that the rotation rate
can tune the coefficients of the corresponding  amplitude equations from regimes where
coherent patterns prevail to regimes of spatio-temporal chaos.
\end{abstract}
 \begin{multicols}{2}

The complex Ginzburg-Landau equation (CGLE)
\begin{equation}\label{onecgl}
 \partial_{t} A = A +
 (1 + i c_{1}) \partial_{x}^{2} A  -
 (1 - i c_3)|A|^2 A ~,
\end{equation}
which describes slow modulations of an amplitude or envelope $A$ 
near a Hopf bifurcation in spatially extended systems,
 has been used extensively both to study nonequilibrium pattern formation
\cite{CH,newell} and as a model system for spatio-temporal chaos
\cite{shraiman,recent}.
The qualitative dynamical behavior of solutions of the CGLE 
depends on the coefficients $c_1$ and $c_3$, which, for a given system,
can be obtained from the underlying equations
by laborious calculations; these have 
been carried out for a small number of systems only \cite{Schoepf}.
For $c_1$ and $c_3$ small, as well as close to the line $c_3\!=\!-c_1$,
the dynamics is close to that found in the relaxational limit
$c_1\!=\!c_3\!=\!0$, whereas for $|c_1|$ and $|c_3|$ large, the CGLE reduces
to the nonlinear Schr\"odinger equation. In recent years, the
complicated and often surprising dynamics that occurs 
away from these limits, has been intensively studied
\cite{shraiman,recent}. In particular, it has become clear that the
CGLE  shows various regimes of spatio-temporal chaos 
when $c_1 c_3\!>\!1$ --- see Fig.\ 1. The
precise nature of the various chaotic regimes, as well as the
existence and nature of the transitions between them,
is still under active investigation in the field of spatio-temporal chaos.

In order to be able to investigate these chaotic regimes {\em
  experimentally}, one  would like to have a
system where the coefficients of the corresponding CGLE can be tuned
in a convenient way through the spatio-temporal chaotic regimes.  When
it was discovered that a forward Hopf bifurcation to a
quasi-one-dimensional wall-mode occurs in rotating Rayleigh-B\'enard
convection in bounded containers \cite{ZhongNing}, it was realized
that the rotation rate might serve to tune the coefficients of the
corresponding CGLE \cite{knob}.  Kuo and Cross therefore performed the
amplitude expansion for such a rotating cylinder of infinite radius
and $\sigma\!=\!6.4$ (corresponding to water), but, unfortunately,
found that the coefficients remain close to the relaxational line
$c_1\!=\! -c_3$ for arbitrary values of the rotation rate $\Omega$
\cite{Kuo}. Our results for the annulus, illustrated by the heavy line for
$\sigma\!=\!6.7$ in Fig.\ 1, confirm this for essentially all Prandtl
numbers larger then 0.2.

\vspace{.4cm}
\epsfysize=6.8cm 
\epsffile{fig1}
FIG 1. 
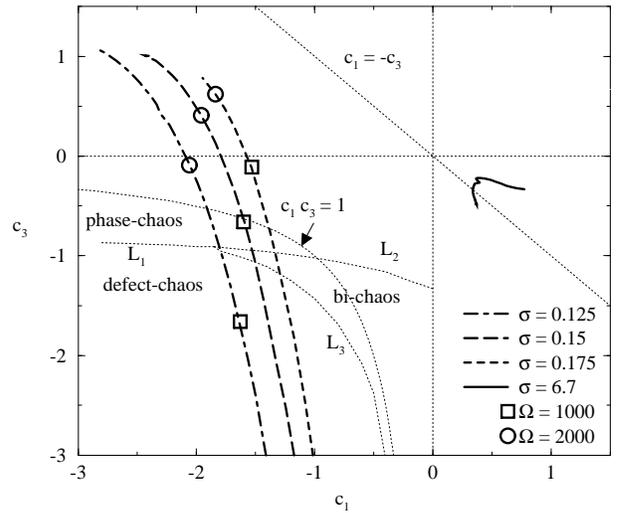
{\small Part of the phase diagram of the CGLE. Since the CGLE
is invariant under $c_1\!\rightarrow\!-c_1$, 
$c_3\!\rightarrow\!-c_3$, so is the phase-diagram; for convenience,
we focus on the left half.
The diagonal line
  $c_1\!=\!-c_3$ indicates the relaxational limit of the 
CGLE, while 
the curve $c_1 c_3 \!=\! 1$ corresponds to the Newell 
criterion \cite{CH};
outside of this hyperbola chaos occurs. The various
spatio-temporal chaotic regimes and the lines $L_1$, $L_2$ and $L_3$
separating them \cite{shraiman,recent} are indicated, and the fat lines
denote the results of our calculations. 
For $\sigma\!=\!6.7$, the coefficients  stay close to the relaxational
limit, similar to \cite{Kuo},
but for $\sigma$ around $0.15$ 
the rotation-rate is able to tune  $c_1$ and $c_3$  from close
to the relaxational limit to deep into the spatio-temporal chaotic
regime. The variation with the dimensionless rotation rate $\Omega$ is
illustrated by the data-points for $\Omega\!=\!1000$ and $\Omega\!=\!2000$.}
\newline

In order to obtain an experimental realization of the CGLE with
coefficients that can be tuned by the rotation rate over a much wider
range, we have investigated convection in rotating annuli for a range
of Prandtl numbers. We take the circumference of the annulus large
enough that the curvature can be neglected.  This geometry differs
from a cylindrical one in that bulk-modes are quasi one-dimensional
and can be described by the same amplitude equations as the
wall-modes, and that there are {\em two} wall-modes, localized near
both side-walls.  The infinite radius limit allows the use of
Euclidian instead of cylindrical coordinates and circumvents the
wavenumber discretization that occurs in a finite container
\cite{Kuo,BG}.  In a system of finite radius, the two wall-modes will
differ slightly, but we will ignore this.  The main result of our
calculations, illustrated in Fig. 1, is that for $0.1\!\lesssim
\!\sigma \! \lesssim 0.2$ {\em traveling waves in a rotating annulus
  are predicted to be an experimental realization of a CGLE whose
  coefficients scan through 
the spatio-temporal chaotic part of the phase diagram
  when the rotation rate is changed}.

The relevant mode in the interesting low  Prandtl number regime is, however,
not the analog of the wall-mode studied by Kuo and Cross \cite{Kuo}. We
have found that the $\sigma$--dependence of the wall mode is weak, and
its coefficients remain close to the 
relaxational limit. However, in an annulus
of finite width, the oscillatory bulk mode identified already long ago
for an infinite container \cite{chandra,clune}, becomes the primary
bifurcation from the non-convective regime for a range of
the rotation rates  and  $0.1\! \lesssim \!\sigma \!
\lesssim 0.2$. It is this bulk mode whose coefficients have the
desired behavior. For even smaller values of $\sigma$, several nearby
branches compete and no simple picture emerges \cite{vanhecke}.
 
Our calculations are based on standard methods  to calculate the
coefficients of the amplitude 
equations (see, e.g.,
\cite{Schoepf}), but are complicated by computational difficulties and the
occurrence of several competing branches of solutions for
$\sigma$. Since a detailed account of the calculations 
and the results is given in \cite{vanhecke},
we confine our presentation here to a description of
the basic setup and a summary of the most important predictions.

The amplitude equations describing the slow modulations of
a right-traveling mode with $x,t$-dependence $e^{i( - k_c x+ \omega_c t)}$ and
amplitude $A$, coupled to a left-traveling
mode $e^{i(  k_c x+\omega_c t)}$ with
amplitude $B$, are \cite{CH}:
\begin{mathletters}\label{coup1dCGLE}
\begin{eqnarray}
\tau_{0} (\partial_{t} + &v_{g} \partial_{x} )A =\varepsilon(1 + i c_{0})A +
\xi_{0}^{2} (1 + i c_{1}) \partial_{x}^{2}A  -\nonumber \\
&g_0 (1 - i c_3)|A|^2 A - g_2 (1 - i c_2) |B|^2 A ~, \label{campa}\\
\tau_{0} (\partial_{t} -& v_{g} \partial_{x} )B =\varepsilon(1 + i c_{0})B+
\xi_{0}^{2} (1 + i c_{1}) \partial_{x}^{2}B  -\nonumber\\
&g_0 (1 - i c_3)|B|^2 B - g_2 (1 - i c_2) |A|^2 B~.  
\end{eqnarray} \end{mathletters}
For Rayleigh-B\'enard convection,  $\varepsilon\!:=\!(R-R_c)/R_c$, with $R$  the Rayleigh number $g \alpha \Delta T d^3/ \kappa \nu$, where
$g$ is the gravitational acceleration,
$\alpha$ the thermal expansion coefficient,
$\Delta T$ the temperature difference between bottom and top plate,
$d$ the height of the layer, 
$\kappa$ the thermal diffusivity
and $\nu$ the kinematic viscosity; 
$R_c$ is the critical
Rayleigh number. These quantities arise in the
equations of motion that describe the fluid-system
in the co-rotating frame, which are the
Navier-Stokes equation with additional
centrifugal and Coriolis force, supplemented by the 
heat equation and the mass conservation law \cite{chandra}.
As usual \cite{Kuo,BG,chandra,clune}, we apply the Boussinesq approximation
and neglect the centrifugal forces.

Below we shall summarize our findings for all the coefficients and
parameters in the amplitude equations. Since we find that $g_2\!>\!g_0$
for the bulk-mode,
the standing waves are suppressed \cite{CH} and the relevant dynamical states
are traveling wave states with, e.g., $A\!\neq \!0$, $B\!=\!0$. Upon
rescaling time, space and the amplitude, Eq.\ (\ref{campa}) then reduces to the
CGLE (\ref{onecgl}) in the frame moving with the group velocity $v_g$.

As length-scale we choose $d$,
so the top and bottom plates
are at $z\!=\!0$ and $z\!=\!1$. We focus here on the case
that the width of the channel is 1; for nearby values of the
width  similar behavior occurs, as detailed in \cite{vanhecke}.
The rotating (infinite radius) annulus
is therefore characterized by two parameters.
The first is the Prandtl number $\sigma\!:=\! \nu/\kappa$.
The second is the dimensionless rotation rate $\Omega\! :=\! 
\Omega_D d^2/\nu$, 
where $\Omega_D$ is the angular velocity.
In a typical experiment,
$\sigma$ is fixed, and the rotation rate
$\Omega$ can be adjusted over a certain range up to values of order $10^4$.
To be able to separate the hydrodynamic equations,
we assume slip boundary conditions 
on the top and bottom plates, as in \cite{Kuo,clune}. 
On the vertical side-walls we apply stick boundary conditions,
which damp the mean-flow that plays a role for low Prandtl number
convection with slip boundary conditions \cite{Siggia}
\begin{mathletters}
\begin{eqnarray}
v_{x} = v_{y} = v_{z} = \partial_{y} \theta = 0
 &\mbox{ on $y$} = 0,&1~, \label{bcy}\\
\partial_{z} v_{x} = \partial_{z} v_{y} = v_{z} = \theta = 0 &\mbox{ on $z$} =
0,&1~, \label{bcz}
\end{eqnarray}
\end{mathletters}
where $\theta$ is the deviation of the temperature from the 
conductive state profile.
It should be noted that our version of the stick boundary conditions
is slightly simpler than those used in \cite{Kuo}.
We have normalized the amplitudes such that $|A|$
represents the ratio of convected to conducted heat; the value
of $g_0$ therefore determines the so-called Nusselt number 
\cite{Kuo}.

{\em Bifurcation structure and linear stability}.  From the equations
of motion we have determined $R_c$, which
is the value of the Rayleigh number where convection sets in, and the
corresponding critical wavenumber $k_c$ and frequency $\omega_c$ as a
function of $\sigma$ and $\Omega$.  
An important
feature of our system is that there exists, in particular for small
$\sigma$'s, a multitude of solutions to the linearized equations, but
only the mode with the {\em lowest} critical Rayleigh number is
relevant.  For a finite cylinder,
these branches are discussed in detail in \cite{BG}.

The relevant features of the bifurcation structure are illustrated in
Fig.\ 2 and can be summarized as follows.
{\em (a)} The linear stability analysis for the stationary modes
does not dependent on the value of $\sigma$ \cite{vanhecke}.
{\em (b)} For all $\sigma\!\gtrsim\!0.2$, the wall-mode is relevant
for
 sufficiently
large rotation rates, while a stationary (non-oscillatory)
mode is relevant for small $\Omega$ --- see Fig. 2(c). For 
$\sigma\!=\!6.7$ 
the crossover between stationary and 
oscillatory mode occurs at $\Omega \approx 27.5$; 
the situation for $\sigma \!=\!6.7$ is representative for the whole range
$\sigma\!\gtrsim\!0.2$. 
{\em  (c)} The bulk-mode exists for values of $\sigma$
that are comparable to where the Hopf bifurcation 
in an infinite layer occurs \cite{chandra,clune}, and approaches this
mode rapidly upon increasing the width of the annulus.
{\em (d)} For $0.1\!\lesssim\!\sigma\!\lesssim\!0.2$ we find a band of
rotation rates $\Omega_{min}\!<\!\Omega\!<\!\Omega_{max}$ for which the
bulk-mode is {\em relevant}. See Fig. 2(c), where
$\Omega_{min}\!=\!140$ and $\Omega_{max}\!=\!5600$ are marked.  
The situation for $\sigma \!=\!0.15$ 
studied below is representative, as  Fig.\ 2(d) for
$\sigma\!=\!0.175$ shows. 
{\em (e)} When $\sigma\!<\!0.1$, the situation becomes quite
convoluted since additional modes become relevant for
some range of $\Omega$'s.

\epsfysize=7.2cm
\vspace{.4cm}
\epsffile{fig2}
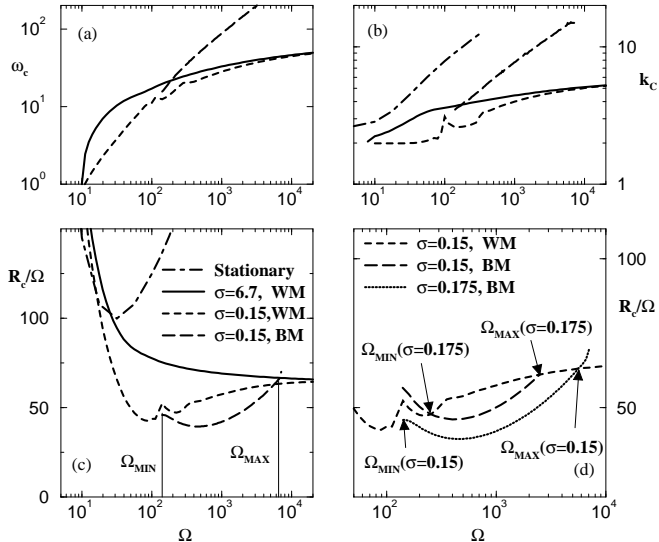
FIG 2. {\small (a)-(c): The linear onset values of bulk- (BM) 
and wall-modes (WM) as a function of
$\Omega$ for $\sigma\! =\!6.7$ and $0.15$.
In (c-d) we have rescaled the critical Rayleigh numbers by 
$\Omega$ to facilitate the comparison
between the various modes.
The stationary bifurcation
is seen to be the primary bifurcation for small rotation rates.
For all $\sigma\!\gtrsim\!0.2$, the wall-mode constitutes the primary mode
for sufficiently large $\Omega$
as illustrated for the case $\sigma \!=\!6.7$.
For $\sigma\!=\!0.15$, the bulk-mode exists for all
$\Omega\!>\!\Omega_{{min}}\!\approx\!140$.
Its critical Rayleigh number is smaller than the
critical Rayleigh number of the wall-mode for
$\Omega_{min}\!<\!\Omega\!<\!\Omega_{max}\!\approx\!5600$.
It is in this range that the coefficients of the CGLE
can be tuned over a wide range.
In (a) and (b) the corresponding critical frequencies and wavenumbers
are plotted.
\small (d): 
For values of the Prandtl number around $0.15$, the critical Rayleigh
numbers of the wall-mode only show a weak dependence
on the Prandtl number, whereas  those
for the bulk-mode strongly depend on $\sigma$.
As a result,
the range of rotation-rates for which the bulk-mode is relevant
strongly depends on $\sigma$;
for $\sigma \!=\!0.175$, the values of $\Omega_{min}$ and $\Omega_{max}$
are approximately $ 200 $ and $2500$, 
while for $\sigma\! =\!0.125$ (not shown)
they are $122$ and $10500$.
}\newline

{\em Amplitude expansion}.  The amount of computer time that is needed
for the calculation of the nonlinear coefficients $g_0,c_3,g_2$ and
$c_2$ of the coupled amplitude equations (\ref{coup1dCGLE}) is
substantial; therefore, we can not scan all system parameters
simultaneously. Instead, we have performed a ``trial and error''
search in the $(\Omega,\sigma)$-space. We shall not exhaust the reader
with the data thus obtained but concentrate on the
wall- and bulk-mode discussed above and to a few values of $\sigma$
that are representative for the various ranges of the Prandtl number.
The dependence of the coefficients $c_1$ and $c_3$ of (\ref{coup1dCGLE})
 for
these two modes is illustrated in Figs.\ 1 and 3.

For the wall-modes, we find that the precise value of 
$\sigma$ is quite irrelevant;
the coefficients $c_1$ and $c_3$ of the amplitude equations
are always near the $c_1\!=\!-c_3$ line, as the full line in  Fig.\ 1
shows for $\sigma\!=\!6.7$.
When $\Omega$ is sufficiently large, 
the wall-modes at the opposite sides of the boundary
are decoupled and we recover 
the result of Kuo and Cross \cite{Kuo}. 
Of course, we have performed extensive searches in 
parameter space to search for more interesting behavior of 
the coefficients $c_1$ and $c_3$,
but for all Prandtl numbers larger than $0.2$,
the wall-mode is relevant, and
the behavior of the coefficients of the amplitude equations (\ref{coup1dCGLE})
is very much like that for $\sigma\!=\! 6.7$.

For all system parameters that
we investigated, $g_2$ remains smaller than 
$g_0$ for the wall-modes, and therefore
the convection occurs in two counter-propagating
traveling waves, that are
localized near the opposite $y$-boundaries \cite{CH}.
The coefficients of the amplitude equations for the wall-modes 
that are relevant for small $\sigma$ and 
$\Omega\!>\!\Omega_{{max}}$, are similar to the coefficients that 
we find for the wall-mode for 
higher Prandtl numbers,  in that $c_1\!\approx\!-c_3$.
In the other small-$\sigma$ regime where the wall-mode is relevant, i.e.,
 $\Omega\!<\!\Omega_{{min}}$, there appears to be
a tiny regime, close to $\Omega_{{min}}$, where the
coefficients $c_1$ and $c_3$ might move away from the dissipative
limit; however, the numerics are not decisive here.

From the point of view of the amplitude equations, the wall-modes
do not have many interesting features, and therefore we will
focus now on the bulk-modes that are relevant 
for small Prandtl numbers and $\Omega_{min}\!<\!\Omega\!<\!\Omega_{max}$.

The coefficients of the amplitude equations (\ref{coup1dCGLE})
for the bulk-mode  are shown
in Fig.\ 3 for $\sigma\! =\!0.15$.
We find that 
the coefficients $c_1$ and $c_3$ can be tuned over a wide range by
the rotation rate. 
For the  system parameters that
we consider, $g_2\!>\!g_0$ [see Fig.\ 3(d)]
and the left- and right-traveling
modes suppress each other. 
The convection patterns thus consist of a juxtaposition of
patches of
left- and right-traveling waves, and after careful adjustments
one may have the convection
exclusively consisting of either a left- or
a right-traveling wave, which warrants a description
with a single CGLE. 
Note that for such a single wave the
value of $c_2$ is immaterial, since terms of the form
$|A|^2B$ or $|B|^2 A$ are zero.

At other values of the Prandtl number $\sigma$ in the range
$0.1\!\lesssim\!\sigma\!\lesssim0.2$, the main effect of a change in
$\sigma$ is through a change in $\Omega_{{min}}$ and
$\Omega_{{max}}$; 
the coefficients of the amplitude equations depend
of $\sigma$,
but this dependence is rather weak, in the sense that $\Omega$
still makes the coefficients vary over a wide range. This is shown in
Fig.\ 1, where the path that the coefficients trace in the parameter
space of the CGLE is shown for three values of $\sigma$.

Can the interesting Prandtl number range
$0.1\!\lesssim\!\sigma\!\lesssim\!0.2$ be accessed experimentally?
Compressed gases have $\sigma\!\gtrsim\!0.7$ and typical liquids have
even larger $\sigma$'s, while liquid mercury and gallium have
$\sigma\!=\!0.025$ and $\sigma\!=\!0.005$. However, Rayleigh-B\'enard
convection in superfluid $^3$He-$^4$He mixtures is known to behave to
a very good approximation as a convecting liquid with a Prandtl number
which can be tuned continuously between 0.02 and about 1
\cite{metcalf}. Convection in rotating cells and flow visualization
have recently become possible for such mixtures \cite{lucas},
and so this system 
provides a unique route to probe the  regimes of
phase-chaos and defect-chaos experimentally, and to compare to
theoretical predictions \cite{shraiman,recent}.

\vspace{.5cm}
\epsfysize=7.5cm
\epsffile{fig3}
FIG 3.
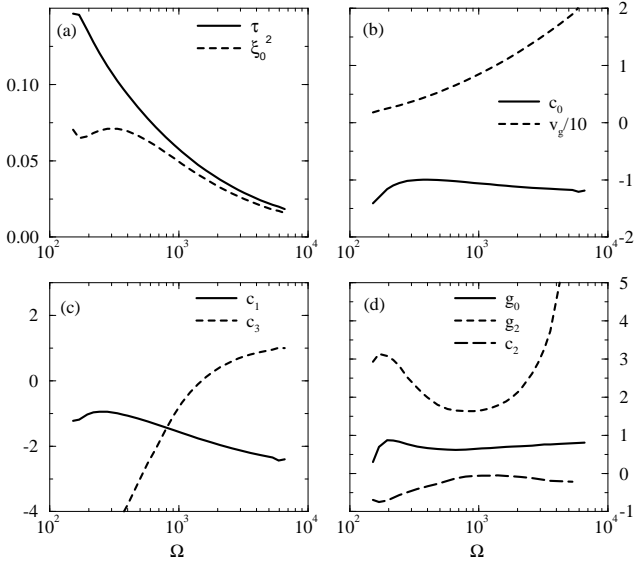
{\small  The coefficients of Eq. (2) for the bulk-mode for
$\sigma\!=\!0.15$.
The coefficients $c_1$ and in 
particular $c_3$ have a strong dependence of the rotation rate (c).
For $\Omega \rightarrow\Omega_{{max}}$, $c_1$ and $c_3$ are close
to the relaxational limit. When the rotation rate is decreased
$c_3$ changes sign, and at $\Omega\approx\!1050$, the Newell criterion
is reached ($c_1 c_3 \!=\!1$). A further decrease of the rotation rate
pushes the coefficients deep into the spatio-temporal chaotic
regime as Fig.\ 1 shows.}
\newline

In summary, we predict that
Rayleigh-B\'enard convection in a large rotating annulus
is an attractive experimental realization of a supercritical CGLE 
with tunable coefficients for a number of reasons:
{\em (i)} The onset of convection can occur either
via a stationary or a Hopf bifurcation;
in the latter case the mode can either consist of 
a {\em single} traveling bulk-wave, or two counter-propagating wall-modes.
The rotation rate can be adjusted to study the competition between 
these states, in analogy to the study of the co-dimension-2 points 
that occur in binary liquid convection \cite{Schoepf}.
{\em (ii)} The quasi one-dimensional geometry of the system warrants a description
in terms of one-dimensional amplitude equations, for both the intrinsic
one-dimensional wall-modes and the intrinsic two-dimensional bulk-mode.
{\em (iii)} The onset of convection occurs via a forward bifurcation; 
for backward bifurcations, like those 
in binary liquids, amplitude equations can at most
give a qualitative description of the patterns. 
{\em (iv)} The underlying basic equations for this system, i.e., the Navier-Stokes
equations, are considerably simpler than the basic equations for
convection in liquid crystals or binary liquids.
For instance, in the latter system, 
it is hard to decide
which aspects of the experimentally observed chaos \cite{paulstc}
can be described by the {\em quintic} CGLE, and which aspects 
are connected to physically relevant effects that are not captured 
in an amplitude description.

We have benefitted from correspondence with  R.\ E. Ecke, M.\ C. Cross
and 
P. Kolodner.

\end{multicols}


\begin{references}
\bibitem{note}
It should be noted that
when the critical Rayleigh numbers of for instance the bulk- and wall-modes
become close, a description in terms of two
coupled amplitude equations is no longer sufficient \cite{CH,Schoepf}.
\end{references}
\end{document}